\documentclass[prl,onecolumn,superscriptaddress,nofootinbib]{revtex4-1}
\usepackage{graphicx}
\usepackage{color}
\usepackage{amsmath}
\usepackage{amsfonts}
\usepackage{mathrsfs}
\begin{document}

\title[Soft Spheres Packing]{Metallurgy of soft spheres with hard core: from BCC to Frank-Kasper phases}

\author{Brigitte Pansu}
\email{brigitte.pansu@lps.u-psud.fr}
\affiliation{Laboratoire de Physique des Solides (CNRS-UMR 8502), B{\^a}t. 510, Universit{\'e} Paris-sud, F 91405 Orsay cedex}

\author{ Jean-Fran\c{c}ois Sadoc }
\affiliation{Laboratoire de Physique des Solides (CNRS-UMR 8502), B{\^a}t. 510, Universit{\'e} Paris-sud, F 91405 Orsay cedex}
\begin{abstract}
To appear in EPJE\\
Understanding how soft particles can fill the space is still an open question. Structures far from classical FCC or BCC phases are now commonly experimentally observed in many different systems. Models based on pair interaction between soft particle are at present much studied in 2D. Pair interaction with two different lengths have been shown to lead to quasicrystalline architectures. It is also the case for a hard core with a square repulsive shoulder potential. In 3D, global approaches have been proposed for instance by minimizing the interface area between the deformed objects in the case of foams or micellar systems or using self-consistent mean field theory in copolymer melts. In this paper we propose to compare a strong van der Waals attraction between spherical hard cores and an elastic energy associated to the deformation of the soft corona. This deformation is measured as the shift between the deformed shell compared to a corona with a perfect spherical symmetry. The two main parameters in this model are: the hard core volume fraction and the weight of the elastic energy compared to the van der Waals one. The elastic energy clearly favours the BCC structure but large van der Waals forces favors Frank and Kasper phases. This result opens a route towards controlling the building of nanoparticle superlattices with complex structures and thus original physical properties.
\end{abstract}

\maketitle
\section{Introduction}

\label{intro}
Understanding how atoms or particles can fill the space is a very old challenge. When particles or atoms behave like hard spheres, the structures that they form are closed-packed structures : face centered cubic (FCC) or hexagonal close-packing (HCP). Other structures like body centered cubic structure (BCC) or Frank-Kasper phases (FK) require more complex interactions \cite{Dzugutov} or anisotropic shapes \cite{Glotzer2012}. Sixty years ago, Frank and Kasper \cite{Frank-Kasper} have investigated many complex alloy structures, constructing packings of polyhedra with large coordination number (Z= 12, 14, 15 or 16) in order to maintain tetrahedral close packing. In the Frank-Kasper (FK) structures, there are at least two types of sites with different environments. This explains why, in atomic systems, these structures are mainly observed in alloys with at least two types of atoms, with some exceptions such as $\beta$-Tungsten (A15 phase) or $\beta$-Uranium ($\sigma$ phase). Since soft particles can adapt their shape to the local geometrical constraints, they are excellent systems to search for FK phases and thus also quasicrystalline structures \cite{Ungar,Talapin}. In these systems, beyond thermodynamic interaction, entropy is suspected to drive new principles of self-organization \cite{Ziherl,Iacovella} and FCC, HCP and BCC structures but also FK phases are expected as for atomic systems. Frank-Kasper phases have already been found in micellar systems \cite{Seddon}, liquid crystals \cite{Ungar,Percec}, star polymers, block-copolymers \cite{Bates2010,Bates2014}.

Micellar systems \cite{Seddon} exhibit mainly two different FK phases: the A15 phase for direct micelles and the C15 phase for inverted micelles. The A15 phase is a cubic phase with Pm3n as symmetry group, a AB$_3$ stoichiometry, 8 spheres per cubic cell and Z=12 (icosahedron) or Z=14 coordination numbers. The C15 Laves phase \cite{Laves} is also cubic with Fd3m as space group, a AB$_2$ stoichiometry, 24 spheres per cubic cell and Z=16 or Z=12 coordination numbers. A tetragonal  FK phase, the $\sigma$ phase (with symmetry P4$_2$/mnm) has been recently discovered in a micellar system \cite{Kim18042017}. FK phases and quasicrystalline phases and their transition sequence have recently been observed in mesophases of one-component giant surfactants \cite{PNAS2016GiantSurf}. A hexagonal FK phase, the C14 phase, that is the hexagonal version of the cubic C15 phase has been discovered in self-assembly of polydisperse populations of charged colloids \cite{Cabane2016}.

The $\sigma$ FK phase has also been reported with dendrimers \cite{Percec} as in polymeric systems \cite{Bates2010,Bates2014}. With 30 spheres per unit cell, it is a good dodecagonal quasicrystal approximant structure primarily reported in numerous metal alloys. Many different Franck and Kasper or Laves phases have been observed in diblock copolymer melts depending how specimens are cooled from the disordered state\cite{Bates2017}. However, in surfactant or block-copolymer self-assembly, the micelles can have different size and the system is thus closer to alloy structure. Combining molecular dynamics simulations have shown that the appearance of these phases are due to tail number variation. Only the dendrimers can be really considered as monodisperse objects building FK phases and dodecagonal quasicrystals.

Metallic or semi-conducting nanoparticles (NP) surrounded by grafted ligands forming a soft corona around them can be considered as soft particles. Moreover they build superlattices very easily \cite{BolesTalapin}. FCC and BCC structures are commonly observed \cite{Landman,Whetten}. Recent experiments have shown that superlattices of dodecanethiol-capped 1.8 nm diameter gold nanocrystals can undergo a non-reversible series of ordered structure transitions at high temperature \cite{Korgel}. Among these phases, several complex pseudo-FK phases have been observed: the cubic NaZn$_{13}$ -type structure (Fm3c), the hexagonal CaCu$_5$ -type structure (P6/mmm). But in this experiment, the structure change is associated to the NP growth and even coalescence leading to various sizes of nanoparticles. More recently, we have reported \cite{HajiwC14} the existence of a Frank-Kasper phase with hexagonal symmetry (MgZn$_2$ type, also labelled C14) in superlattices of monodisperse hydrophobically-coated gold particles at room temperature obtained from suspensions in various solvents. These different results show than even if the core is solid the corona softness can be sufficient to induce complex packing architectures.

 The final structure built by soft sphere systems results from a delicate balance between different energies but the driving mechanisms can be very different: electrostatic interaction, steric repulsion, elastic deformation free energy, van der Waals attraction, surface tension...There is certainly no universal answer to the question of what structure for soft spheres packing. In copolymers, the entropic stretching usually favors a BCC structure and one can suspect that the elastic deformation free energy also. This is the first ingredient of the model. But more complex architectures require another ingredient. In this paper, the second ingredient will be van der Waals force. The model that we propose thus considers an assembly of soft objects with a rigid core and a soft shell, regularly located at the main sites of crystalline structures. The model balances attractive vdW interaction between the cores with a soft elastic energy which attempts to mimic the behavior of the shell. The shell elastic energy is related to its deformation and thus to the local environment of each particle. This environment is naturally described by the Voronoi cell at the considered site. The simplest model assume that the soft objects fill the Voronoi cell around each site. This elastic energy is measured by the shift between the deformed shell and a corona with a perfect spherical symmetry. This approach is reminiscent of previous studies on block copolymers \cite{PhysRevLett.72.936,GRASON2006}. The two main parameters of the model are the volume fraction occupied by the hard cores and the ratio between the elastic energy parameter and the van der Waals one.  By comparing different structures, a quantitative phase diagram involving different phases (BCC, A15, C14, C15, $\sigma$) is deduced from this model. Whereas the elastic energy favors the BCC structure, the C14 Frank and Kasper phases is induced by large enough van der Waals forces. The A15 or $\sigma$ phase are close in energy and appear in between. Comparing with experimental systems requires a physical analysis of the origin of the elastic energy in each system that is often a difficult point. Nevertheless this model suggests both new design principles for superlattice formation in nanoparticle assemblies and further opens the door to extend this heuristic approach to more complex particle systems.

\section {What energies are involved in the model for the determination of the final structure? }

 Soft nanoparticles dispersed in solvent interact via various potentials. Considering the effect of the van der Waals forces between the cores is certainly pertinent for different systems. For metallic nanoparticles dispersed in oil or in water, the van der Waals attraction is notably large due to large electronic fluctuations. This is also probably the case for diblock copolymers with large Flory Huggins parameter.
 In our model, the cores are assumed to be rigid and spherical. The van der Waals interaction between two cores with diameter $D_0$ distant of $r$ is given by \cite{Israelachvili}:

\begin{equation}
V_{vdW}(r,D_0)=- \frac{A}{12}\left( \frac{D_0^2}{r^2-D_0^2}+\frac{D_0^2}{r^2}+2Ln\left(\frac{r^2-D_0^2}{r^2}\right)\right).
\end{equation}

The van der Waals energy per particle can be computed by summing the van der Waals energy between pairs of particles, neglecting any collective effect. For each site, the sum must include not only the first neighbors but all the particles in a large isotropic volume of the structure surrounding the considered site that are required for a proper convergence of the sum. For a 24\% core volume fraction, at least neighboring crystallographic cells must be taken into account to have less than 3\% error on the final value. When there are different sites in the crystallographic cell, the final value will be the mean value on all the sites. Let us introduce a first important parameter, that is the volume fraction occupied by the rigid cores $\varphi$. For each structure, the cell parameter depends only on $\varphi$ and on the core diameter $D_0$ or the core radius $r_c=D_0/2$. This energy will be very sensitive to the distance between the cores that is directly correlated to $\varphi$. But the final van der Waals energy depends only on $\varphi$ and is independent on the core diameter since $r=D_0 f(\varphi)$. In the FCC structure, the distance between the cores is the largest one, this structure is thus not favored at all by the van der Waals attraction. Since it is also not favored by the elastic interaction, the FCC structure will be ignored in the following.
	
\begin{figure}
\center
\resizebox{0.3\textwidth}{!}{%
\includegraphics{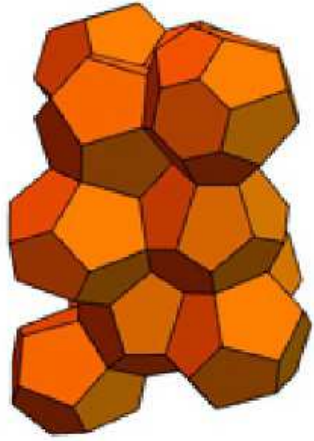}
}
\caption{ The Voronoi cells of the twelve crystallographic sites in the unit cell of the Frank-Kasper C14 structure. In this example there are 8 cells with 12 pentagonal faces and 4 cells with 12 pentagonal faces and 4 hexagonal faces. }
\label{fig:VoroC14}
\end{figure}

 The elastic shell energy that will balance the van der Waals attraction between the cores is related to its deformation of the soft shell and thus to the local environment of each particle. This environment is naturally described by the Voronoi cell at the considered site. For example, Figure \ref{fig:VoroC14} shows different Voronoi cells in the case of the C14 structure. The Voronoi cell is not necessarily unique in the different structures and varies a lot between the different structures even if they have to follow generic rules in the case of FK phases. To compare the different structures, we shall consider that the mean Voronoi cell volume is the same for all the considered structures that are compared. Table~\ref{tab:volVoro} gives the normalized volumes of the different Voronoi cells for several structures. Variations of about 10\% are observed but these values are not sufficient to estimate the elastic free energy due to the difference of the Voronoi cells with a spherical shape of unit volume. The main goal of this paper is to estimate properly the elastic free energy associated to this deformation in order to balance the attractive van der Waals energy.

\begin{table}
\small
\caption{ Volume of the Voronoi cell(s)for all the studied structures (F-K phase and BCC): C14, C15, A15, $\sigma$, Z and BCC. The normalized volumes (the mean cell volume is 1) are given for the different coordination shells: Z12, Z14, Z15 and Z16. If two different volumes appear for the same coordination type, they are indicated as $Zx_{a~\rm{or}~b}$. The quadratic deviation for these volumes is given under the mean coordination $\overline{z}$. All geometric figures are drawn using Mathematica.}
\label{tab:volVoro}

\begin{tabular}{ l c c c c c c }

\hline\noalign{\smallskip}
Struct. 	&C14 				& C15 			& A15 		&$\sigma$		& $Z$		& BCC \\
										
\hline\noalign{\smallskip}
 nb of 	&12					&24					&8				&30					&7				&2 \\
sites	& 				& 					& 				& 					& 				& \\
\hline\noalign{\smallskip}
 		Z12$_a$& 				& 			& 		& 		& 	& \\
 	nb&2					&16					&2				&2					&3				& \\
cell vol.						&0.935			&0.9288			&0.9765		&0.9188			&0.9044		&		\\
\hline\noalign{\smallskip}
		Z12$_b$& 				& 			& 		& 		& 	& \\
	 nb	&6					& 					& 				& 8					& 				& \\
cell vol.						&0.9283			& 					& 				& 0.9524		& 				&		\\
\hline\noalign{\smallskip}
 	Z14$_a$& 				& 			& 		& 		& 	& \\
	 nb	& 					& 					& 6				& 8					& 2				& 2 \\
cell vol.						& 					& 					& 1.0078	& 1.0317		& 1.02042	&	1	\\
\hline\noalign{\smallskip}
 Z14$_b$	& 				& 			& 		& 		& 	& \\
	 nb 	& 					& 					& 				& 8					& 				& \\
cell vol.						& 					& 					& 				& 1.009			& 				&	 	\\
\hline\noalign{\smallskip}
 	Z15& 				& 			& 		& 		& 	& \\
	 nb			& 					& 					& 				& 8					& 2 			& \\
cell vol.						& 					& 					& 				& 1.0707		& 1.1228 	&	 	\\
\hline\noalign{\smallskip}
 Z16& 				& 			& 		& 		& 	& \\
 nb 			&4 					& 8					& 				& 					& 			& \\
	cell vol.					&1.1397			& 1.1423		& 				& 					& 			 	&	 	\\
\hline\noalign{\smallskip}
$\overline{z}$&13.333		& 13.333	&13.5				& 13.467 	& 13.429 			& 14 \\

$\sqrt{\delta}$&0.0988	& 0.1006	& 0.0135		& 0.0472	& 	0.091		 	&	0 	\\
\noalign{\smallskip}\hline
\end{tabular}
\end{table}

\begin{figure}
\resizebox{0.45\textwidth}{!}{%
\includegraphics{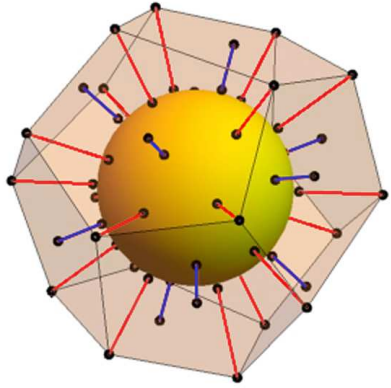}
}
\caption{\label{fig:ligandVoro} Spokes radiating from a gold core toward surface of a Voronoi cell in a possible Frank-Kasper phase (here 12-face cell in the A15 structure). The longest spokes (in red) correspond to the more extended region; they are directed towards the Voronoi cell vertices. The shortest ones (in blue)  correspond to the more compressed region; they are directed towards the centers of the faces.}
\end{figure}

 To estimate the free energy linked to the Voronoi cell deformation, we consider the distance $\ell$ from the core surface to the Voronoi surface. For a perfect spherical particle this distance is constant and equal to $\ell_0$. For a deformed particle, this distance is no longer equal to $\ell_0$ and the length shift $\delta \ell=\ell -\ell_0$ varies around the core as shown in Figure \ref{fig:ligandVoro}. The elastic deformation energy is estimated by

\begin{equation}
U_{el}=\frac{1}{2} k \left< {(\delta \ell)}^2\right>
\end{equation}
with
\begin{equation}
k=\frac{K}{(\ell_0)^2}
\end{equation}

 where the mean value is taken all around the core. $K$ has the dimension of an energy relating the elastic deformation to the relative length variation $\delta \ell/\ell_0$. The distance $\ell_0$ can be related to $\varphi$ and to the rigid core radius ${r_c}$:
 \begin{equation}
 \frac{4\pi}{3}r_c^3= \varphi \left[\frac{4\pi}{3}(r_c+\ell_0)^3\right].
\end{equation}

that is:
\begin{equation}
 \ell_0= r_c \left({\varphi}^{-1/3} -1\right).
\end{equation}

To compute the mean value of $\delta \ell ^2$, we consider a large number $N_s$ of spokes radiating from a core towards the surface of the Voronoi cell surrounding this core. The positions of the spokes on the core have been chosen using semi-regular polyhedra or a phyllotactic organization on the sphere \cite{Sadoc2013} in order to adjust more easily their number $N_s$ and their uniform distribution. The mean elastic energy per spoke for a given Voronoi cell takes the form of a sum running on all the spokes (indexed by $i$) contained in the Voronoi cell:

\begin{equation}
 U_{el}=\frac{1}{2 N_L} K \Sigma_{i=1}^{N_l} \sqrt{\frac{||{\bf r}_{i,k}-{\bf c}_{i,k}||^2}{(\ell_0)^2}}
\end{equation}
where ${\bf c}_{i,k}$ is the vectorial position of the grafting point of the spoke $i$ in the Voronoi cell $k$ whereas ${\bf r}_{i,k}$ is the vectorial position of the free end of the spoke.
 $N_s$ has been chosen close to 100. Since the shape of the Voronoi cell is not isotropic, this energy is sensitive to the orientation of the core. Therefore the elastic energy is calculated for several randomly chosen orientations of the set of coordinates ${\bf c}_{i,k}$ of all spoke origins on all cores in the crystallographic unit cell. As the obtained energy slightly depends on this choice of orientations, the final elastic energy is therefore the average value on different orientations.

The total energy which drives the relative stability of the different phases depends on the van der Waals energy per particle and on the elastic energy of the soft shell per particle. Let us introduce the ratio $R=K/A$ of the elastic constant $K$ over the Hamaker constant $A$, $R$ has no unit. This ratio is an important parameter in the balance between the shell elastic energy and the van der Waals attraction between the cores. A large $R$ will favor the shell elastic part of the total energy whereas a low value of $R$ will favor the van der Waals attraction between the cores. The other important parameter that drives both the distance between the cores and the shell thickness reference is the core volume fraction $\varphi$.

\begin{figure}
\center
\resizebox{0.45\textwidth}{!}{%
\includegraphics{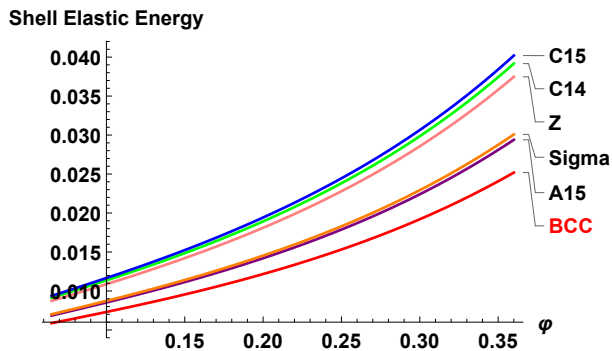}
 }
\caption{ The normalized elastic energy for $K=1$, for BCC, C14, C15, A15, $\sigma$ and Z phases as function of the volume fraction $\varphi$. }
\label{fig:ElastEner}
\end{figure}

Figure \ref{fig:ElastEner} shows the elastic energy (with $K=1$) as a function of $\varphi$ for six different phases. The elastic energy increases with increasing core volume fraction for all structures. When comparing the different energies, the BCC structure is clearly the most favorable one. The A15 and $\sigma$ phases are close in energy but less favorable than the BCC phase. The C14, C15 and Z phase are also close in energy but with higher values. It has already been shown that the entropic stretching energy evaluated in the block copolymer framework generically favors BCC over Frank Kasper phases like A15 \cite{Grason2003}. Predicting the relative behavior of the elastic energy for the different Frank Kasper phase is not so easy.  It is not related to the close packing volume fraction (BCC:68\%, A15=52.3\%, C14=55.5\%). It appears that how energies are ordered is related to the deviations of the Voronoi cell volumes compared to the mean volume in the different phases. Large mean coordination number also seems to increase the energy. Nevertheless it is important to notice that the cubic C15 phase is less favorable than the homologous hexagonal phase C14 in terms of ligand elastic energy that was quite unexpected. This model is therefore an efficient way to get a quantitative comparison of the elastic energy of different phases.

\begin{figure}
\resizebox{0.45\textwidth}{!}{%
\includegraphics{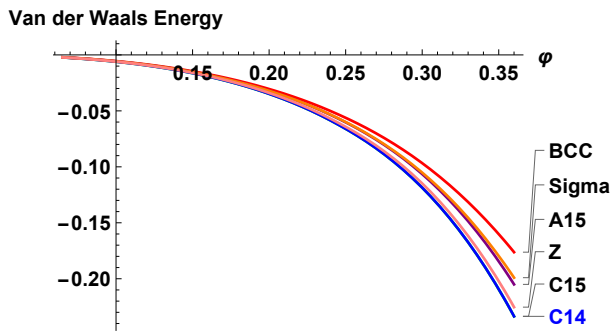}
 }
\caption{The normalized van der Waals energy ($A=1$), for BCC, C14, C15, A15, $\sigma$ and Z phases as function of the $\varphi$.}
\label{fig:VdW}
\end{figure}

Figure~\ref{fig:VdW} shows the van der Waals energy per particle as a function of $\varphi$ for six different phases and $A=1$. From the van der Waals point of view, the comparison leads to a totally reverse conclusion compared to the elastic deformation energy : the C14 and C15 phases are clearly favored as well as the Z phase. The BCC phase is the less favorable one. The $\sigma$ phase and the A15 phase have similar intermediate behavior.
The total energy per particle combines the van der Waals energy and the total elastic energy: it depends on $\varphi$, $A$ and $K$. But the phase diagram depends only on $\varphi$ and on $R=K/A$.
\begin{equation}
 U_{tot}(\varphi)= U_{vdW}(\varphi, A)+U_{el}(\varphi, K).
\end{equation}

\begin{figure}
\resizebox{0.45\textwidth}{!}{%
\includegraphics{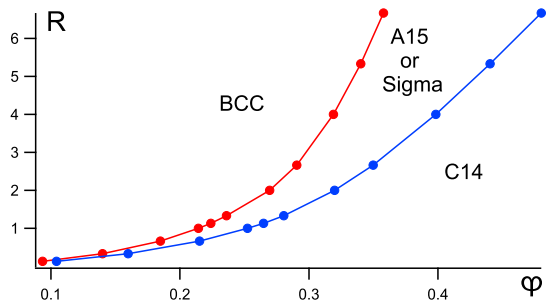}
 }
\caption{ Phase diagram of soft particles as a function of the core volume fraction $\varphi$ and of the ratio $R$ of the elastic constant of the soft shell over the Hamaker constant that drives the van der Waals attraction. The shell elasticity favors the BCC structure whereas the van der Waals attraction between the hard spherical cores favors the C14 structure. The A15 or Sigma structures appear in between.}
\label{fig:Diagram}
\end{figure}

Figure ~\ref{fig:Diagram} shows the phase diagram of soft particles as a function of the core volume fraction $\varphi$ and of the ratio $R$ of the elastic constant of the soft shell over the Hamaker constant. The shell elasticity (large $R$) favors the BCC structure whereas the van der Waals attraction between the hard spherical cores (small $R$) favors the C14 structure. The A15 structure appears in between. But the $\sigma$ structure has a very close energy and thus the A15 and $\sigma$ structures cannot be distinguished in this model. The
energy of the C15 structure which is the cubic version of the hexagonal C14 structure is always larger than the energy of the C14 structure.

	 	This elastic model remains a crude approach of the ligand response upon shell deformation. Even if our approach to quantify the elastic deformation of the shell is rough, it enables a quantitative comparison of the different phases and thus a prediction of phase diagrams for systems where the van der Waals attraction is expected to be large. This is certainly the case for metallic core particles. The Hamaker constant of gold is expected to be quite high due to the metallic state of gold even in small cores and is usually estimated to 75 $k_BT$ for gold in oil. This model thus explains well the phase diagram of small gold nanoparticles with a core diameter $D_0$ close to 2 nm. They self-assemble either in the C14 structure for short ligands (hexane-thiol) \cite{HajiwC14} or to BCC phase with longer ligands (dodecane-thiol) \cite{Schmittemulsion}. The repulsion from the ligand shell is not really known whereas the structure of the ligands in certainly an important key to understand the various structures \cite{Travesset2017}. Unfortunately there is no theory available for interacting olimeric ligands as those used in this system. Nevertheless the model that we propose is in good agreement with the observed structures. For the 2nm-core gold nanoparticles, the C14 structure (and not the C15 structure) is observed for a core volume fraction close to 24\% and the BCC structure for a core volume fraction close to 18\%. This is coherent with the phase diagram of Figure ~\ref{fig:Diagram} if the ratio $R$ is close to 0.8. This would correspond to a constant $K$ close to 60 $k_BT$. The A15 or $\sigma$ phase is predicted for intermediate ligand length. Experiments are at present performed to test this prediction but to distinguish clearly between a C14 phase and a $\sigma$ phase requires high quality superlattices and experiments are carried out to grow them.

	Other models are proposed to explain and predicts Frank and Kasper phases or even quasicrystalline phase for soft sphere self-assembly. The model proposed in this paper does not take into account any interfacial energy. The core is considered as rigid ans thus the interface between the core and the shell remains constant. This is not the case for many soft systems as block copolymers. In such systems, this interfacial energy is known to play an important role. In a previous study \cite{Grason2005} of melts composed of block copolymers with multiply-branched architecture, it has been shown that the A15 cubic phase is stabilized over the BCC phase by the tendency of the AB interfaces to conform to the polyhedral environment of the Voronoi cell of the micelle lattice. In other soft matter systems like in foams or in micellar systems, the energy is mainly related to surface tension; for instance the surface tension of the films around the bubbles. The surface area of the Voronoi cells will thus play an important role in these systems \cite{Weaire1996}. In this model, no surface tension between the shells is assumed.

	 In the Frank and Kasper phases or Laves phase, all the crystallographic sites are not equivalent and Voronoi cells have different volumes. Nevertheless, there is a modification of the Voronoi construction (Laguerre construction \cite{SadocJullien}) which allows to have a weighting factor controlling cell size. It is then possible to adjust weight in order to have all cell volumes equal. We have done this construction with the phases presented above and it appears that the total elastic energy is only very lightly modified using this construction. The elastic energy is mainly due to fluctuations of the crown thickness which could be expected to be lowered when all cells have an equal volume, but cells are more irregular in that case and then there is no gain for the elastic energy.

In the same way, this model can easily be adapted to mixtures of nanoobjects with different size or nature. Indeed, since the crystallographic sites are not equivalent, alloys like in metallic crystals are expected to stabilize these phases. In the model presented in this paper, the polydispersity has been neglected whereas it is suspected to stabilize Frank-Kasper phases and extend their range in the phase diagram. It should be included for a better description of the self-assembly of soft objects.
	
	A very important point in all the experiments is the way the superlattices are grown. The BCC structure as well as the C14 structure are clearly equilibrium phase for gold nanoparticles self assembly. But all the possible structures are close in energy. In block copolymers, it is clear that different processing routes drive assembly into a variety of low-dimensional phases more typical of metal alloys \cite{Bates2017}. For the gold nanoparticles, different self-assembly mechanisms \cite{phdthesisBorn} can be used with various kinetics of the agglomeration process. As in diblock copolymers, different process routes are suspected to achieve different structures. Controlling the superlattice growth is thus essential to explore the phase diagram and ensure that the observed phase is really the thermodynamically stable one.
\section{ Conclusion}	
	
	 There is certainly no universal answer to the question of what structure for soft spheres packing. In this paper, we focus our attention on soft spheres with hard core interacting through strong van der Waals attraction. The final structure for this soft sphere systems is assumed to result from a delicate balance between the van der Waals attraction and the elastic ligand deformation energy. A quantitative estimation of the elastic deformation of the soft corona is proposed. The van der Waals energy per particle as well as the mean value of the elastic energy has been computed for different structures, classical ones (BCC) but also more complex structures as various Frank and Kasper or Laves phases. The different structures can be compared depending on the hard core volume fraction and on the ratio between the elastic constant and the Hamaker constant. This model is well adapted to hydrophobically coated gold nanoparticles. It explains why the hexagonal C14 phase and not the close C15 cubic phase has been observed for short ligands whereas the structure is BCC for longer ligands. It also predicts how other phases like the $\sigma$ phase or the A15 phase could be expected and thus can help to choose the good parameters in order to obtain the desired structure and search for quasicrystalline phases.


%
%

\section{Acknowledgement}
The authors thank Gregory Grason for fruitful discussion on the elastic energy and the different behavior from this point of view of the C14 and C15 phases. The first discussion has been completed during the stay of one of us (JFS) at Aspen Center for Physics during a workshop organized by Gregory Grason (National Science Foundation grant PHY-1607611).

\section{Authors contributions}
All the authors were involved in the preparation of the manuscript.
All the authors have read and approved the final manuscript.
%



\section*{References}

\begin{thebibliography}{10}



\bibitem{Dzugutov}
 M. Dzugutov (1993). \emph{Phys. Rev. Lett.}  \textbf{70}, 2924\textendash2927.

\bibitem{Glotzer2012}
  P. F. Damasceno, M. Engel and S. C. Glotzer (2012). \emph{Science}  \textbf{337}, 453\textendash457.

\bibitem{Frank-Kasper}
F. C. Frank and J. S. Kasper (1958). \emph{Acta Crystallogr.}  \textbf{11}, 184\textendash190.

\bibitem{Ungar}
G. Ungar \& X. Zeng (2005). \emph{Soft Matter}  \textbf{1}, 95\textendash106.

\bibitem{Talapin}
D. V. Talapin, E. V. Shevchenko, M. I. Bodnarchuk1, X. Ye, J. Chen \& C. B. Murray (2009). \emph{Nature}  \textbf{461}, 964\textendash967.

\bibitem{Ziherl}
P. Ziherl \& R. D. Kamien (2001). \emph{J. Phys. Chem. B}  \textbf{105}, 10147\textendash10158.

\bibitem{Iacovella}
C. R. Iacovella, A. S. Keys \& S. C. Glotzer (2011). \emph{Proc. Natl. Acad.}  \textbf{108}, 20935\textendash20940.

\bibitem{Seddon}
J. Seddon \& R. Templer (1995), in Polymorphism of Lipid-Water
Systems, ed. R. Lipowsky and E. Sackmann  \emph{Elsevier Science}, 97\textendash160.

\bibitem{Percec}
G. Ungar, Y. Liu, X. Zeng, V. Percec \& W.-D. Cho (2003). \emph{Science}  \textbf{299}, 1208\textendash1211.

\bibitem{Bates2010}
S. Lee, M. J. Bluemle \& F. S. Bates (2010).  \emph{Science}  \textbf{330}, 349\textendash353.

\bibitem{Bates2014}
L. Sangwoo, C. Leighton \& F. S. Bates (2014). \emph{Proc. Natl. Acad.}  \textbf{111}, 17723\textendash17731.

\bibitem{Laves}
F. Laves (1949). in ``Crystal Chemistry: Structure of Metals, Metalloids and their Compounds''. \emph{London: Butterworth}.

\bibitem{Kim18042017}
S. A. Kim, K.-J. Jeong, A. Yethiraj \& M. K. Mahanthappa (2017). \emph{Proc. Natl. Acad.}  \textbf{114}, 4072\textendash4077.

 \bibitem{PNAS2016GiantSurf} K. Yue, M. Huang, R.L. Marson, J. He, J. Huang, Z. Zhou,
J. Wang, C. Liu, X. Yan, K. Wu et al.(2016). \emph{PNAS} \textbf{113}, 14195


\bibitem{Cabane2016}
B. Cabane, J. Li, F. Artzner, R. Botet, C. Labbez, G. Bareigts,
M. Sztucki \& L. Goehring (2016). \emph{Phys. Rev. Lett.}  \textbf{116}, 208001.

\bibitem{Bates2017}
K. Kim, M. W. Schulze, A. Arora, R. M. Lewis, M. A. Hillmyer,
K. D. Dorfman \& F. S. Bates (2017). \emph{Science}  \textbf{356}, 520\textendash523.


\bibitem{BolesTalapin}
M. A. Boles, M. Engel \& D. V. Talapin (2016). \emph{Chemical Reviews}  \textbf{116}, 11220\textendash11289.

\bibitem{Landman}
U. Landman \& W. D. Luedtke (2004). \emph{Faraday Discuss.}  \textbf{125}, 1\textendash22.

\bibitem{Whetten}
R. L. Whetten, M. N. Shafigullin, J. T. Khoury, T. G. Schaaff,
I. Vezmar, M. M. Alvarez \& A. Wilkinson (1999). \emph{Acc. Chem. Res.}  \textbf{32}, 397\textendash406.

\bibitem{Korgel}
B.W. Goodfellow, M. R. Rasch, C. M. Hessel, R. N. Patel, D.-M.
Smilgies and B. A. Korgel (2013). \emph{Nano Lett.}  \textbf{13}, 5710\textendash5714.

\bibitem{HajiwC14}
S. Hajiw, B. Pansu \& J.-F. Sadoc (2011). \emph{ACS Nano.}  \textbf{9}, 8116\textendash8121.

\bibitem{PhysRevLett.72.936}
P.D. Olmsted, S.T. Milner (1994). \emph{Phys. Rev. Lett.} \textbf{72}, 936 

\bibitem{GRASON2006}
G.M. Grason (2006). \emph{Physics Reports} \textbf{433}, 1

\bibitem{Israelachvili}
J. N. Israelachvili ed.(2011), ``Intermolecular and Surface Forces'', \emph{ Academic Press, San Diego}  3rd edn.

\bibitem{Sadoc2013} J. F. Sadoc, J. Charvolin \& N. Rivier (2013). \emph{Journal of Physics A: Mathematical and Theoretical}  \textbf{46}, 295202.


\bibitem{Grason2003}
G.M. Grason, B.A. DiDonna, R.D. Kamien (2003). \emph{Phys. Rev. Lett.} \textbf{91}, 058304

\bibitem{Schmittemulsion}
J. Schmitt, S. Hajiw, A. Lecchi, J. Degrouard, A. Salonen M. Impéror-Clerc \& B. Pansu (2016). \emph{The Journal of Physical Chemistry B}  \textbf{120}, 5759\textendash5766.


\bibitem{Travesset2017} A. Travesset (2017). \emph{ACS Nano} \emph{11}, 5375

\bibitem{Grason2005}G.M. Grason, R.D. Kamien (2005), \emph{Phys. Rev. E} \emph{71}, 051801




\bibitem{Weaire1996}
D. Weaire. Ed., ``The Kelvin problem: foam structures of minimal surface area'' (1996). \emph{London; Bristol, PA: Taylor and Francis}.

\bibitem{SadocJullien}
Sadoc, J. F., Jullien, R., Rivier, N.(2003). \emph{Eur. Phys. J. B} \textbf{33}, 355


\bibitem{phdthesisBorn}
P. G. Born (2011). PhD thesis, \emph{ Universität des Saarlandes, Saarbrücken}.
%
%
\end{thebibliography}
\end{document}